\documentstyle[aps,prb,epsf]{revtex}
\epsfclipon

\begin{document}

\twocolumn[\hsize\textwidth\columnwidth\hsize\csname 
@twocolumnfalse\endcsname

\title{Conductivity of a 2DEG in Si/SiGe heterostructure near metal-
insulator transition: role of the short and long range scattering 
potential}

\author{E.~B.~Olshanetsky$^1$, V.~Renard$^{2,3}$, Z.~D.~Kvon$^1$, 
J.~C.~Portal$^{2,3,4}$,\\
N.~J.~Woods$^5$, J.~Zhang$^6$, 
J.~J~Harris$^6$}

\address{$^1$ Institute of Semiconductor Physics, Novosibirsk 630090, Russia;}
\address{$^2$ GHMF, MPI-FKF/CNRS, BP-166, F-38042, Grenoble, Cedex 9, France;}
\address{$^3$ INSA-Toulouse, 31077, Cedex 4, France;}
\address{$^4$ Institut Universitaire de France, Toulouse, France;}
\address{$^5$ Imperial College of Science, London WC7 2BW, UK ;}
\address{$^6$ University College, London WC 15 75, UK;}
\maketitle

\begin{abstract}
  We report the observation of a metal-insulator transition (MIT) in a two-
dimensional electron gas (2DEG) in a Si/SiGe heterostructure at zero
magnetic field. On going through the MIT we observe the corresponding
evolution of the magnetic field induced transition between the insulating
phase and the quantum Hall (QH) liquid state in the QH regime. Similar to
the previous reports for a GaAs sample, we find that the critical magnetic
field needed to produce the transition becomes zero at the critical electron
density corresponding to the zero field MIT. The temperature dependence of
the conductivity in a metallic-like state at zero field is compared with the
theory of the interaction corrections at intermediate and ballistic regimes
$k_{B}T\tau/\hbar\geq1$. The theory yields a good fit for the linear part of
the curve. However the slope of that part of $\sigma_{xx}(T)$ is about two
times smaller than that reported in other 2D systems with similar values of
$r_s$. At the same time, the recent theory of magnetoresistance due to
electron-electron interaction in the case of arbitrary $k_{B}T\tau/\hbar$,
smooth disorder and classically strong fields does not seem to be quite
adequate for the description of the parabolic magnetoresistance observed in
our samples. We attribute these results to the fact that neither of these
theories deals with the whole scattering potential in a sample but leaves
either its long range or its short range component out of consideration.

\end{abstract}

]
\bigskip
\narrowtext

\section{Introduction}

  According to the well established one-parameter scaling theory of
conductivity \cite{scaling} any two dimensional electron system can exist
only in insulating state and no MIT should be possible. Thus, it came quite
as a surprise when an apparent metallic state and an MIT were observed first
in Si MOSFETs \cite{firstMIT} and, later, in several other 2D systems, such
as p-Si/SiGe, n- and p-GaAs/AlGaAs etc \cite{MITreview}. The fact that the
MITs have always been observed in systems with a large parameter $r_s$ (the
ratio of the interaction energy to the kinetic energy) is an indication that
electron-electron interaction plays an important role in the MIT. While it
still remains unclear whether the observed MIT is a genuine transition and
whether a zero temperature metallic state can really exist at certain levels
of interaction and disorder, its discovery has stimulated an extensive
experimental and theoretical work in the area. In particular, a considerable
progress has been made \cite{Zala} in our understanding of the temperature
dependent corrections to the conductivity from electron-electron interaction
not only in the diffusive regime $k_{B}T\tau/\hbar\ll1$, as previously, but
also in the ballistic and the intermediate regimes $k_{B}T\tau/\hbar\geq1$,
typical for systems showing a MIT. The theoretical results
\cite{Zala,Zala2,Zala3} have been reported to give a good description of the
conductance temperature dependence and magneto- resistance in parallel
magnetic field in several types of 2D systems on the metallic side of the
apparent MIT, \cite{Dolgopolov,Hole_Hole,Pudalov,Brunthaler,Vitkalov}. It
should, however, be noted, that the theory \cite{Zala,Zala2,Zala3} relies on
an approximation where the scattering potential is treated as point-like
scatterers. While appropriate for Si MOSFETs, this approximation is
definitely not true for high mobility heterostructures where the impurities
are separated from the 2DEG by an undoped spacer and the scattering
potential is predominantly long range. Therefore, a certain caution should
be used when applying the results of Ref.~\onlinecite{Zala,Zala2,Zala3} for
the description of the experimental data in high mobility heterostructures. 

  Till recently, in the context of the new approach \cite{Zala} no explicit
discussion has been made of the interaction contribution to the
magnetoresistance in a transverse magnetic field. The point has been
recently addressed in Ref.~\onlinecite{Gornyi} where the interaction
contribution to the magnetoresistance in a classically strong transverse
field is calculated for a smooth disorder and for arbitrary
$k_{B}T\tau/\hbar$. Soon afterwards the results of Ref.~\onlinecite{Gornyi}
have been tested in an experiment on magnetoresistance in a GaAs/AlGaAs
heterostructure with a 2DEG and a good agreement with theory \cite{Gornyi}
has been found \cite{Proskur}. 

  In the present work we attempt an experimental investigation of these
questions in a Si/SiGe heterostructure with a 2DEG. Despite the abundance of
information on the MIT in Si MOSFETS and different other types of
heterostructures, so far there has been no observation of an MIT in a
n-Si/SiGe. This may in part be explained by the difficulty in fabricating a
gated n-Si/SiGe structure with a stable and controllable behavior. At the
same time an experimental study of these questions in a n-Si/SiGe is
expected to be of some interest. Such a study would provide an opportunity
to compare the electron transport properties of a Si MOSFET and a Si/SiGe
heterostructure in the vicinity of the MIT and in the ``metallic'' regime.
This comparison may be instructive as these two silicon structures have very
similar electron energy spectrum, but differ in the structure of the
disorder potential, which is predominantly short range in Si MOSFETs but has
an important long range component in a Si/SiGe heterostructure. In this work
we report the first observation of the MIT in a Si/SiGe heterostructure with
a 2D electron gas and perform an analysis of the magnetoresistance data and
the conductance temperature dependence on the metallic side of the MIT using
the results of Ref.~\onlinecite{Zala} and Ref.~\onlinecite{Gornyi}. The
theory \cite{Zala} produces a good fit for the linear part of the
conductance themperature dependence. However the value of the interaction
constant $F_{0}^{\sigma}$ serving as the only fitting parameter turns out to
be about two times smaller than that reported for Si MOSFETs with similar
values of $r_s$, \cite{Dolgopolov}. Also, in contrast to \cite{Proskur} we
find a rather big divergence between our magnetoresistance data and the
behavior predicted in Ref.~\onlinecite{Gornyi}. We attribute all these
results to the possibility that in the intermediate and ballistic regimes
the case of a mixed mechanism of scattering, i.e. the situation where both
the short and the long range scattering are equally important for electron
transport, cannot be properly described by a model that ignores either one
of the other component of disorder. 

\section{Experimental procedures}  
 
  The n-type modulation doped Si/Si$_{1-x}$Ge$_{x}$ samples used in this
study were grown by MBE \cite{material}. On a high resistivity p-type Si
substrate a 1.4$\mu$m graded Si/Si$_{1-x}$Ge$_x$ layer, in which the Ge
content increases linearly from 5 to 35\%, was grown, followed by 0.5$\mu$m
of constant composition Si$_{0.7}$Ge$_{0.3}$ alloy. These layers produced a
strain-relaxed but low-dislocation-density surface onto which the tensilely
strained Si channel, 110 $\AA$ thick, was deposited, followed by an undoped
Si/Si$_{1-x}$Ge$_x$ spacer layer, a doped Si/Si$_{1-x}$Ge$_x$ supply layer
and finally undoped Si/Si$_{1-x}$Ge$_x$ and Si cap layers. The thickness and
As doping levels in the supply layer were varied slightly between the
samples. To measure the magnetoresistivity of the samples, the layers were
photolithographically processed into 50 $\mu$m wide Hall bars with the
distance between the four voltage probes on each side 100, 250 and 100
$\mu$m. The mesa pattern for the Hall bars was etched by reactive ion
etching to define the channel geometry. The ohmic contacts to the Hall bars
were prepared by ion doping of phosphorus.
 
  The magnetotransport properties of the structures were measured at 0.4-4.2
K in magnetic fields up to 15 T using a superconducting solenoid and He$_3$
and VTI cryostats. The data was acquired using standard a.c. lock-in
techniques with a frequency around 10 Hz. Driving currents through the Hall
bar of 0.1 $\mu$A were chosen to avoid the heating effects. All our samples
behaved in a more or less similar way. In what follows we discuss the
results obtained on a sample with the electron density
$n_{s}=(3.53-6.23)\times 10^{11}$cm$^{-2}$ depending on the prior
illumination and the highest mobility $\mu\approx 6\times 10^{4}$
cm$^{2}$/Vs.
   
\section{Results and Analysis}

\begin{figure}
\centerline{\epsfxsize3in\epsfbox{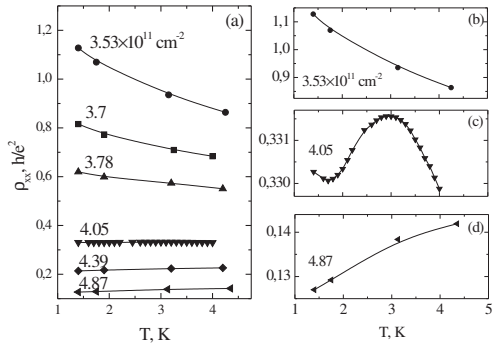}\bigskip}
\caption{a-resistivity versus temperature for different electron densities;
b,c,d - some of the traces from Fig.1a given in more detail}
\label{fig1}
\end{figure}

  As has been mentioned in the introduction the nature of the so called MIT
remains a topic of debate. Since all the experiments are necessarily
conducted at finite temperatures it is impossible to say whether the
metallic state observed in these experiments would persist down to zero
temperature or whether it will ultimately evolve into an insulating state.
Despite a large number of articles on the MIT in various types of 2D
semiconductor systems, so far there has been no report of a MIT in a Si/SiGe
heterostructure with a 2DEG. The MIT is typically observed by measuring the
temperature dependence of a sample resistance for different carrier
densities which are usually controlled by voltage applied to the
electrostatic top gate. Reportedly, it has been found difficult to fabricate
a n-Si/SiGe heterostructure with a gate that does not leak. The leakage from
the gate to the 2DEG results in a sample instability that makes it
unsuitable for experiment. In the present work, after several unsuccessful
attempts to fabricate an operational gated n-Si/SiGe structure, we decided
to perform our study using structures without gate. Among the several
samples that we have tried there were some that, upon cooling down to low
temperatures in the dark were found in high resistance low electron density
states. The samples were also found to be very sensitive to illumination.
Exposing them to a specially selected dose of red LED radiation permitted us
to fine-tune the sample resistance with a good precision. The sample states
of different resistivity thus obtained were characterized by a high
stability yielding reproducible dependences of resistance versus temperature
and magnetic field over a considerable, up to several hours, period of time.
Shown in Fig.~\ref{fig1} is a series of resistance versus temperature traces
for different electron densities obtained using the procedure described
above. Similar to the previously reported MITs in other 2D semiconductor
systems \cite{MITreview} we observe a transition from the insulator to
metallic-like behavior as the electron density is varied from $3.53\times
10^{11}$cm$^{-2}$ to $4.87\times 10^{11}$cm$^{-2}$ . To our knowledge this
is the first observation of this kind in a Si/SiGe heterostructure with a
2DEG. In Fig.~\ref{fig1},b,c,d some of the curves are given in more detail.
With our method of varying the electron density and a rather limited
temperature range used it is difficult to determine with a high precision
the "separatrix", i.e. the curve corresponding to the transition from the
metallic to insulator regime. Nevertheless we consider the curve shown in
Fig.~\ref{fig1}c to be close to that transition point. There are two things
about this curve that are worth mentioning. Firstly, the resistivity of the
sample corresponding to the separatrix is about $0.3h/e^2$ which is about
ten times lower than that reported for Si MOSFETs \cite{MITreview}.
Secondly, the peculiar non-monotonic temperature dependence corresponding to
our separatrix is also quite unusual. At present we have no explanation for
these observations.
 
\begin{figure}
\centerline{\epsfxsize3in\epsfbox{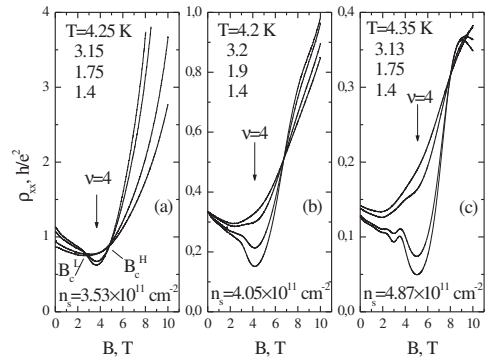}\bigskip}
\caption{The magnetoresistance traces taken at different temperatures 
corresponding to some of the sample states shown in Fig.1}
\label{fig2}
\end{figure}

  Before the first observation of the MIT in Si MOSFET \cite{firstMIT} it
has been widely accepted that the electrons confined to two dimensions have
an insulating ground state at zero magnetic field and zero temperature. On
the other hand the application of a strong perpendicular magnetic field
changes the situation, resulting in a transition form the insulating state
to a QH liquid state. This magnetic-field-driven transition from the
insulator to the QH liquid state has attracted much attention \cite{QH1} and
some interesting hypothesis have been proposed for its explanation
\cite{QH2,QH3}. Later, it was suggested \cite{Tsui} that there possibly
exists a link between the MIT transition observed at zero magnetic field and
the finite magnetic field transitions studied previously. In the present
work we have made a similar observation in our samples. Fig.~\ref{fig2}
shows the magnetoresistance traces taken at different temperatures
corresponding to some of the zero field sample states presented in
Fig.~\ref{fig1}. For any given magnetic field it is the sign of the
resistance temperature dependence that is the indicator of whether the state
of the sample at this particular field is insulating or metallic. Let us
start with a sample in an insulating state at zero field. In
Fig.~\ref{fig2}a we see that for $n_{s}=3.53\times 10^{11}$cm$^{-2}$ the
zero field insulating state persists up to $B^{L}_{c}\approx2.8$T where the
first insulator-to-QH liquid transition occurs. This QH liquid state
corresponding to filling factor $\nu=4$ is then maintained until the QH
liquid-to- insulator transition turns the sample again into insulator at
$B^{H}_{c}\approx4.8$T . According to Fig.~\ref{fig1}, increasing the
electron density from $n_{s}=3.53\times 10^{11}$cm$^{-2}$ brings the sample
closer to the zero field MIT. It is interesting to note, that in magnetic
field this results in a corresponding gradual decrease of $B^{L}_{c}$ until,
finally, at the carrier density $n_{s}=4.05\times 10^{11}$cm$^{-2}$
appropriate for the zero field MIT, $B^{L}_{c}$ becomes exactly zero,
Fig.~\ref{fig2}b. On further increasing the carrier density, the zero field
metallic state is maintained at finite field and the first
magnetic-field-driven transition to occur is thus a QH liquid-to-insulator
transition at higher fields, Fig.~\ref{fig2}c. As the authors of
Ref.~\onlinecite{Tsui}, we find the behavior described above to be an
indication that both the finite- and zero-field transitions are likely to
share a common physical origin.

  Up to date there is still no theoretical description available for the
behavior of a 2D system in the vicinity of the MIT, i.e. when
$\rho_{xx}\approx h/e^2$ , and all the discussions of the MIT are,
therefore, necessarily qualitative in character. On the contrary there is
now a number of comprehensive theoretical models for a situation when
$\rho_{xx}\ll h/e^2$. These models offer predictions concerning the
corrections to the zero temperature Drude conductivity
$\sigma_{D}=e^{2}n\tau/m^{\ast}$ under various physical conditions. We now
turn to the discussion of these quantum corrections in the limit
$\rho_{xx}\ll h/e^2$, i.e. when our samples show metallic behavior. 

  On the whole, there are two distinct types of corrections to conductivity.
First, there is the weak localization correction calculated in the
approximation of non-interacting electrons and describing the variation of
conductivity due to the electron coherent scattering by the random potential
\cite{band4}:
\begin{equation} 
\label{Eq1}
\Delta\sigma_{xx}^{wl}=\alpha p(e^{2}/\pi h)ln(k_{B}T\tau/\hbar),
\end{equation}
  where it is assumed that the phase breaking time $\tau_{\phi}$ varies as
$T^{-p}$ . The amplitude $\alpha$ is expected to be 1 for normal scattering
(-0.5 for pure spin-orbit scattering and 0 for scattering on magnetic
impurities). In Si samples only the Coulomb scattering is present. The above
expression for weak localization correction to conductivity remains valid
for all 2D systems with $\rho_{xx}\ll h/e^2$ . The situation is more
complicated when the interaction between electrons is taken into account.
The first theory to consider the interaction corrections to conductivity was
developed for the diffusive regime $k_{B}T\tau/\hbar<1$ i.e. for a situation
where the electron experiences multiple scattering by impurities during its
life time \cite{Altshuler}. Of the two components of the interaction
correction in this regime, the correction in the Cooper channel is
negligible when $T_{c}\ll T$ , where $T_c$ is the critical temperature for
superconductor transition, and the remaining correction in the Diffusion
channel has the form:
\begin{equation} 
\label{Eq2}
\Delta\sigma_{xx}^{ee}=\frac{e^2}{2\pi^{2}\hbar}
ln\left[\frac{k_{B}T\tau}{\hbar}\right]
\left[1+3\left[1-\frac{ln(1+F_{0}^{\sigma})}{F_{0}^{\sigma}}\right]\right]
\end{equation}
  where $F_{0}^{\sigma}$ is the interaction constant in the triplet channel.
This constant is negative if the electron-electron interaction tends to
align the electron spins and its absolute value increases with the
interaction strength characterized by parameter $r_s$ . Note that the sign
of this logarithmic in temperature correction depends on the absolute value
of $F_{0}^{\sigma}$ . It is believed that in the limit of very strong
interaction the constant $F_{0}^{\sigma}$ approaches -1 leading to the
Stoner instability with all electron spins parallel in the absence of
magnetic field.
 
  Recently, a theory has been proposed \cite{Zala}, that along with the
diffusive regime covers the intermediate and ballistic regimes
$k_{B}T\tau/\hbar\geq 1$. It is shown that the interaction corrections in
all these regimes are a consequence of the same physical process, namely the
coherent electron scattering on the modulated density of other electrons
(Friedel oscillations) caused by an impurity with a short range potential.
The quantum interaction correction in this theory can be expressed as:
\begin{equation} 
\label{Eq3}
\Delta\sigma_{xx}^{ee}=\delta\sigma_C+15\delta\sigma_T
\end{equation}
where 
$$\delta\sigma_C=\frac{e^2}{\pi\hbar}\frac{k_{B}T\tau}{\hbar}\left[1-\frac{3}{8}
f(k_{B}T\tau/\hbar)\right]-
\frac{e^2}{2\pi^2\hbar}ln\left[\frac{1}{k_{B}T\tau/\hbar}\right];$$
is the charge channel correction and
$$\delta\sigma_T=\frac{F_{0}^{\sigma}}{\left[1+F_{0}^{\sigma}\right]}
\frac{e^2}{\pi\hbar}\frac{k_{B}T\tau}{\hbar}
\left[1-\frac{3}{8}t(k_{B}T\tau/\hbar;F_{0}^{\sigma})\right]$$
$$-\left[1-\frac{1}{F_{0}^{\sigma}}ln(1+F_{0}^{\sigma})\right]
\frac{e^2}{2\pi^2\hbar}ln\left[\frac{1}{k_{B}T\tau/\hbar}\right];$$
  is the correction in the triplet channel. We also take into account the
two fold valley degeneracy at (100) Si surface that changes the triplet term
prefactor from 3 to 15. The explicit expressions for the functions
$f(k_{B}T\tau/\hbar)$ and $t(k_{B}T\tau/\hbar;F_{0}^{\sigma})$ are given in
Ref.~\onlinecite{Zala}. In the diffusive limit Eqs.(\ref{Eq3}) reproduce the
well known logarithmic temperature correction Eq.(\ref{Eq2}). In the
intermediate and ballistic regimes Eqs.(\ref{Eq3}) result in a linear with
temperature correction to conductivity with the slope and the sign
determined by the absolute value of the interaction constant
$F_{0}^{\sigma}$. For small $F_{0}^{\sigma}$ the triplet channel
contribution is small and the temperature dependence, governed mostly by the
correction in the charge channel, is insulating, whereas for higher
$F_{0}^{\sigma}$ values the negative triplet term increases and the behavior
becomes metallic. Also, since the triplet term prefactor increases with the
valley degeneracy, systems with a higher valley degeneracy show metallic
behavior for lower values of $F_{0}^{\sigma}$ and hence for lower
interaction strength than systems with lower or no valley degeneracy. 

\begin{figure}
\centerline{\epsfxsize3in\epsfbox{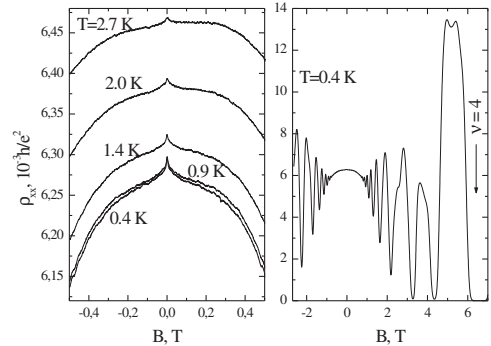}\bigskip}
\caption{a- magnetoresistance in a metallic-like state of the sample at 
different temperatures; b-magnetoresistance shown in a wider field 
range at $T=0.4K$}
\label{fig3}
\end{figure}

\begin{figure}
\centerline{\epsfxsize3in\epsfbox{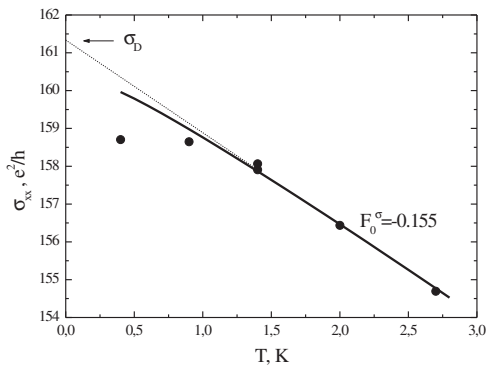}\bigskip}
\caption{Conductivity versus temperature at zero field obtained 
from the magnetoresistance traces in Fig.3a. The thick solid 
line is the sum of the localisation
term and the interaction correction, \cite{Zala} 
with $F_{0}^{\sigma}=-0.155$.}
\label{fig4}
\end{figure}

  Fig.~\ref{fig3} shows the typical magnetoresistance traces obtained at
different temperatures in our samples after the electron concentration had
been saturated with a maximum LED illumination. The corresponding electron
mobility is $\mu=61800$cm$^{2}$/Vs and the electron density $n_s=6.23\times
10^{11}$cm$^{-2}$. In Fig.~\ref{fig4} we plot the zero field conductivity
values form Fig.~\ref{fig3}a versus temperature. In line with the earlier
report on a high mobility n-Si/SiGe structute,\cite{Kawaji} one can see in
Fig.~\ref{fig4} that the sample behavior is markedly metallic. The
temperature dependence appears to be linear for $T\geq1.25$K and saturates
at lower temperatures. By extrapolating the linear part of the curve to
$T=0$ we find the Drude conductivity and the corresponding momentum
relaxation time $\tau=6.8\times 10^{-12}$s. Hence, we get
$k_{B}T\tau/\hbar=0.89\times T$ which means that in the experimental
temperature range $0.4-2.7$K the sample operates in the intermediate and
ballistic regimes. The solid line in Fig.~\ref{fig4} is a theoretical fit
given by the sum of the weak localisation term (Eq.(\ref{Eq1})) and the
interaction correction (Eqs.(\ref{Eq3})) with the only fitting parameter
being the interaction constant $F_{0}^{\sigma}=-0.155$. The weal
localisation term prefactor $\alpha p=1.5$ used in the fit is obtained from
the analysis of the magnetoresistance data in low fields. 

  It is impossible to compare the calculated value of $F_{0}^{\sigma}$ with
the experimental parameter $r_s$ which is $\sim6.7$ for our sample since the
functional relation between $F_{0}^{\sigma}$ and $r_s$ is unknown for
$r_s>1$. It is, however, interesting to note that systematically higher
values of $F_{0}^{\sigma}$ have been reported for various other types of 2D
systems and, in particular, for Si MOSFETs with $r_s$ values close to ours.
For example, $F_{0}^{\sigma}\approx-0.27$ has been found in an n-channel Si
MOSFET \cite{Dolgopolov} and in a p-type GaAs/AlGaAs heterostructure
\cite{Hole_Hole} with $r_{s}\approx 7$. The difference in the
$F_{0}^{\sigma}$ values obtained in our Si/SiGe heterostructure and in an
n-channel Si MOSFET \cite{Dolgopolov} is, in our opinion, the consequence of
a difference in the structure of the scattering potential in these two
silicon systems. The linear-in-$T$ term of Ref.~\onlinecite{Zala} arises
only due to the scattering on a short range impurity potential (in a purely
smooth disorder this term is exponentially suppressed). In a Si MOSFET with
a predominantly short range scattering potential the results of
Ref.~\onlinecite{Zala} are, thus, quite appropriate. In our samples the
single particle scattering time $\tau_q$ obtained from the magnitude of the
Shubnikov-de Haas oscillations is about 6 times smaller than the momentum
relaxation time $\tau$ indicating that apparently both types of disorder are
present. Therefore the prefactor in front of the linear term would be
reduced in our case as compared to the corresponding expression in
Ref.~\onlinecite{Zala} and, hence, a lower value of $F_{0}^{\sigma}$ would
be obtained from fitting the theory \cite{Zala} to experiment which, under
these conditions, is unlikely to reflect the actual interaction intensity in
the sample. This effect is expected to be even more pronounced in such high
mobility structures as n- and p- GaAs/AlGaAs and, in our opinion, a certain
caution should be used when comparing experiment with theory \cite{Zala} in
these cases.

  On the whole the theory \cite{Zala} produces a good fit for the
experimental curve at temperatures $T\geq1.25$K. On the lower temperature
side the dependence $\sigma_{xx}(T)$ tends to saturate, in contrast to the
theoretical prediction. Similar departures from theory have been observed in
all our samples and also by other authors in a Si 2D system different from
ours \cite{Pudalov}. A non-zero valley splitting and a strong sample
specific inter-valley scattering have been suggested as a possible
explanation of the weakening of the $\sigma_{xx}(T)$ dependence. At present
there is no theory available that takes inter-valley scattering into
account. 

  Let us now analyze the transverse magnetoresistance in our samples,
Fig.~\ref{fig3}a. In the absence of any quantum corrections to the
conductivity one has: $\sigma_{xx}^{0}(B)=\sigma_{D}/\left[1+(\mu
B)^{2}\right]$, $\sigma_{xy}^{0}(B)=-\mu B \sigma_{xx}^{0}(B)$,
$\rho_{xy}^{0}(B)=-\mu B/\sigma_{D}$ and
$\rho_{xx}^{0}(B)=1/\sigma_{D}=const$, i.e. the resistance of the sample is
independent of magnetic field. If, on the other hand, there are some small
corrections to the conductivity tensor components
$\sigma_{xx}(B)=\sigma_{xx}^{0}(B)+\Delta\sigma_{xx}$ and
$\sigma_{xy}(B)=\sigma_{xy}^{0}(B)+\Delta\sigma_{xy}$, then, converting the
conductivity tensor into the resistivity tensor, one gets:
\begin{equation} 
\label{Eq4}
\rho_{xx}(B)=\rho_{D}+\rho_{D}^{2}\left[\left((\mu B)^2-1\right)
\Delta\sigma_{xx}+2\mu B\Delta\sigma_{xy}\right]
\end{equation}
  In general there will be contributions to the magnetoresistance both from
the weak localization and from the interaction. Weak localization gives rise
to magnetic-field-dependent corrections to $\sigma_{xx}^{0}(B)$ and
$\sigma_{xy}^{0}(B)$ but these are suppressed and can be neglected in
magnetic fields higher than $B_{tr}\approx\hbar/2el_{e}^{2}$ which is $\sim
1mT$ for our sample. The situation with the interaction corrections to the
magnetoresistance is more complicated. In the diffusive limit
$k_{B}T\tau/\hbar\ll1$ it has been established that the longitudinal
conductivity correction $\Delta\sigma_{xx}^{ee}$, given by Eq.(\ref{Eq2}),
remains unchanged at low and classically strong magnetic fields, while, at
the same time, the Hall conductivity is not affected by interaction:
$\Delta\sigma_{xy}^{ee}\equiv 0$ \cite{Altshuler}. As can be seen from
Eq.(4) for $\omega_c\tau>1$ this will lead to a parabolic negative
magnetoresistance:
\begin{equation} 
\label{Eq5}
\rho_{xx}(B)=\rho_{D}+\rho_{D}^{2}(\mu B)^2
\Delta\sigma_{xx}^{ee}(T)
\end{equation}
  with $\Delta\sigma_{xx}^{ee}(T)$ given by Eq.(\ref{Eq2}). Finally, at
still higher fields $B_Z\geq kT/g\mu_{B}$ a positive magnetoresistance
develops caused by the Zeeman effect on the interaction correction
$\Delta\sigma_{xx}^{ee}$, \cite{Altshuler}. 

  In contrast to the diffusive regime, the behavior of magnetoresistance is
less well understood in the intermediate and ballistic regimes
$k_{B}T\tau/\hbar\geq1$. The theory of interaction corrections at arbitrary
relation between temperature and elastic mean-free time \cite{Zala} makes
certain predictions concerning the corrections to the Hall coefficient
\cite{Zala2} and magnetoresistance in a parallel field \cite{Zala3} but
leaves magnetoresistance in a perpendicular magnetic field out of
consideration. Thus it remains unclear whether the corrections to
conductivity obtained in this theory for zero field would persist in finite
magnetic field and give rise to a magnetoresistance in a manner similar to
the diffusive regime, Eq.(\ref{Eq5}), or whether they will be modified or
suppressed by magnetic field. As has been mentioned above, the results of
the theory \cite{Zala} are obtained in the approximation where the
impurities are treated as point-like scatterers. This is justified at zero
magnetic field where a long-range potential reduces the backscattering and
suppresses the interaction corrections in the ballistic regime. The
situation, however, changes in a strong magnetic field which increases the
probability of an electron to return back and, thus, restores the
interaction correction even in the case of a purely smooth disorder.
Recently, a theory \cite{Gornyi} has been proposed that considers the
interaction in the ballistic and intermediate regime in classically strong
fields $\omega_c \tau>1$. The theory \cite{Gornyi} takes into account only
the long-range component of the scattering potential and shows that in this
case the interaction will also lead to a parabolic negative
magnetoresistance described by Eq.(\ref{Eq5}) with a distinct expression for
$\Delta\sigma_{xx}^{ee}(T)$. However, as will be shown below, while
justified in the case of a predominantly long range scattering potential,
the exclusion of the short range component of disorder from consideration
may be inappropriate in a more general case of a mixed mechanism of
scattering. Finally, it might be well to point out that unlike the diffusive
regime the conductivity correction $\Delta\sigma_{xx}^{ee}(T)$ extracted
from magnetoresistance in the ballistic regime \cite{Gornyi} has nothing to
do with the correction at zero field (the latter being exponentially
suppressed in the case of a smooth disorder). 

\begin{figure}
\centerline{\epsfxsize3in\epsfbox{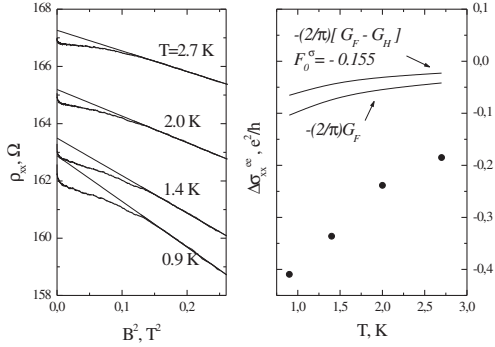}\bigskip}
\caption{a-magnetoresistance data from Fig.3a plotted versus 
$B^2$; b-the black circles are 
$\Delta\sigma_{xx}^{ee}(T)$ obtained 
using Eq.(\ref{Eq5}) from the slope of
the straight part of the magnetoresistance 
traces in Fig.5a; the solid curves are 
theory \cite{Gornyi}, (see text).}
\label{fig5}
\end{figure}
    
  Up to date there have been several experimental studies of the
interaction-related transverse magnetoresistance in the intermediate
magnetic field range where the weak localization is already suppressed while
the Zeeman effect on the interaction correction has not yet set in. First to
mention are the experiments \cite{Paalanen,Minkov,Choi,Thornton} where the
observed temperature dependent negative magnetoresistance was compared to
the theoretical prediction for the diffusive regime, Eq.(5). In one of these
works \cite{Thornton} the samples used were n-Si/SiGe heterostructures very
similar to ours. However with the exception of Ref.~\onlinecite{Minkov}
these experiments have been performed on high mobility structures in the
ballistic regime where the validity of Eq.(5) for the description of the
magnetoresistance is questionable. Recently an experimental work has been
reported \cite{Proskur} whose aim was to study the magnetoresistance caused
by electron-electron interactions in the intermediate regime in a
n-GaAs/AlGaAs heterostructure with long range fluctuation potential and to
compare the results with the theoretical prediction \cite{Gornyi}. A good
agreement between the theory and experiment has been reported.
  
  Similar to \cite{Proskur} in our case we have a relatively high mobility
2D electron gas in the intermediate and ballistic regimes. The important
role of the long range scattering potential is evidenced by the fact that
the transport time is significantly longer than the quantum single-particle
time. The sharp change of the magnetoresistance observed in Fig.~\ref{fig3}a
in low magnetic field is caused by the WL effect suppressed at higher
fields. Then follows a relatively flat region, which, according to
\cite{Gornyi}, is supposed to correspond to the suppression of
backscattering by the long-range potential in low fields. This effect is
more pronounced for higher temperatures. At still higher field the
backscattering is restored by the increasing magnetic field and a parabolic
magnetoresistance sets in. In our sample the condition $\omega_c \tau= 1$ is
satisfied at $B=0.16$T and so the parabolic magnetoresistance in
Fig.~\ref{fig3}a is observed in classically strong fields. On the other
hand, except for $T=0.4$K, the magnetoresistance traces in Fig.~\ref{fig3}a
are measured at fields well below $B_{Z}$[T]$=0.74\times T$[K] at which the
Zeeman effect on the interaction correction should become important. That
means that the parabolic magnetoresistance in Fig.~\ref{fig3}a is observed
in the magnetic field region where the results of Ref.~\onlinecite{Gornyi}
are expected to be applicable. In Fig.~\ref{fig5}a we plot the resistivity
from Fig.~\ref{fig3}a as a function of $B^2$. Shown in Fig.~\ref{fig5}b are
$\Delta\sigma_{xx}^{ee}(T)$ obtained using Eq.(\ref{Eq5}) from the slope of
the straight part of the magnetoresistance traces in Fig.~\ref{fig5}a. The
theoretical curves plotted in Fig.~\ref{fig5}b are
\begin{equation} 
\label{Eq6}
\Delta\sigma_{xx}^{ee}=-\frac{2}{\pi}G_{F}(k_{B}T\tau/\hbar)
\end{equation}
and
$$\Delta\sigma_{xx}^{ee}=-\frac{2}{\pi}
\left[G_{F}(k_{B}T\tau/\hbar)-G_{H}(k_{B}T\tau/\hbar; F_{0}^{\sigma})\right]$$
  where $G_{F}(k_{B}T\tau/\hbar)$ is the exchange contribution to the
interaction correction to conductivity in classically strong magnetic fields
and in the case of a long range scattering potential.
$G_{H}(k_{B}T\tau/\hbar; F_{0}^{\sigma})$ is the triplet Hartree term. The
explicit expressions for these functions are given in
Ref.~\onlinecite{Gornyi}. For plotting in Fig.~\ref{fig5}b we take these
functions in the appropriate for our experimental situation limit $\kappa\gg
k_F$, where $k_F$ is the Fermi wave vector and $\kappa=4\pi e^2 \nu$ is the
inverse screening length. The Hartree term is plotted for
$F_{0}^{\sigma}=-0.155$, the interaction constant obtained from the fit of
the zero-field resistance temperature dependence to theory \cite{Zala}. As
can be seen there is a rather big divergence between the experimental points
and the theoretical curves in Fig.~\ref{fig5}b. The exchange correction
independent of the interaction constant lies closer to the experiment.
Taking into account the Hartree contribution with a sign opposite to the
exchange term increases the discrepancy, which becomes even greater if it is
multiplied by a factor $5$ corresponding to the valley degeneracy, (compare
Eqs.(\ref{Eq3})). In our opinion the explanation why in contrast to
Ref.~\onlinecite{Proskur} we don't find a good agreement with
Ref.~\onlinecite{Gornyi} in our experiment may lie with the fact that theory
\cite{Gornyi} was developed with only the long-range component of the
disorder potential taken into account. The structures studied in
Ref.~\onlinecite{Proskur} were of a higher mobility than ours and with a
more important role of the long-range scattering potential compared to that
of the point-like scatterers. Probably in that case the approximation made
in Ref.~\onlinecite{Gornyi} is adequate, whereas in our samples the
short-range scattering can not be completely neglected and an extension of
theory \cite{Gornyi} to the case of a mixed disorder is needed to explain
our results.

\section{Conclusions}

  To conclude, we have observed for the first time a MIT in a Si/SiGe
heterostructure with a 2DEG. We have also observed the evolution of the
magnetic-field-driven transition from an insulator to a QH liquid state as
our system is step by step driven through the MIT by gradually increasing
the electron density. Similar to Ref.~\onlinecite{Tsui} we find that the
magnetic field corresponding to the insulator-to-QH liquid state transition
becomes zero at the critical electron density appropriate for the MIT in
zero field. On the metallic side of the MIT we find that the theory
\cite{Zala} of the interaction corrections to conductivity for an arbitrary
relation between temperature and the momentum relaxation time provides a
good description of the linear part of the conductivity temperature
dependence at zero field but with the interaction constant $F_{0}^{\sigma}$
about two times smaller than in Si MOSFETs with a similar value of $r_s$.
The recently developed theory \cite{Gornyi} of the interaction contribution
to magnetoresistance at classically strong magnetic fields and for arbitrary
$k_{B}T\tau/\hbar$ does not seem to be appropriate for the description of
the parabolic negative magnetoresistance observed in our system. The theory
\cite{Gornyi} was built in the assumption that there is only a long-range
scattering potential. The importance of a short-range disorder in our
samples may be the reason for the observed discrepancy with the theory. Both
these theories \cite{Zala,Gornyi} neglect either the long or the short range
component of disorder and for that reason do not seem to be quite adequate
for the description of the case of a mixed disorder. 

  We are grateful to I.V. Gornyi and A.D. Mirlin for helpful discussion.
This work was supported by PICS-RFBR (N 1577.), RFBR (N 02-02-16516), NATO
Linkage (N CLG.978991), INTAS (N01-0014), programs "Physics and Technology
of Nanostructures" of the Russian ministry of Industry and Science and "Low
dimensional quantum structures" of RAS.

\end{document}